\documentclass[twocolumn,pra,a4paper,floatfix,preprintnumbers,amsmath,amssymb]{revtex4}

\usepackage{amsmath}
\usepackage{graphicx}

\usepackage{graphicx}
\usepackage{dcolumn}
\usepackage{bm}

\newcommand{\ud}{\mathrm{d}}

\newcommand{\be}{\begin{equation}}
\newcommand{\ee}{\end{equation}}

\begin{document}

\title{Sub Shot-Noise interferometric phase sensitivity \\
with Beryllium ions Schr\"odinger Cat States}

\author{Luca Pezz\'e and Augusto Smerzi}
\affiliation{CNR-INFM BEC Center and Dipartimento di Fisica - Universit\`a di Trento, I-38050 Povo, Italy }


\begin{abstract}
Interferometry with NOON quantum states can provide
unbiased phase estimation with a sensitivity scaling as $\Delta \theta \sim 1/N_T$ given 
a prior knowledge that the true phase shift $\theta$ lies in the interval 
$-\pi \leq \theta \leq \pi$. The protocol requires a total of $N_T = 2^{p}-1$
particles (unequally) distributed among $p$ independent measurements and overcomes 
basic difficulties present in previously proposed approaches.
We demonstrate the possibility to obtain a phase sensitivity beating the classical 
shot-noise limit using published probabilities retrieved experimentally for the creation 
of Schr\"odinger cat quantum states containing up to $N=6$ beryllium ions.
\end{abstract}

\maketitle

\textbf{Introduction.} 
Interferometry plays a central role in the development of basic science and new technologies.
Its main goal is to estimate phase shifts generated by the interaction of the interferometer with 
some external perturbation in domains spanning from micro-scales, as in the measurement of 
Casimir forces, to cosmic-scales, as in the detection of gravitational waves.
The precise estimation of phases is limited by two quite different sources of noise. 
Classical noise can be created by micro-seismic geological activities, 
temperature fluctuations, poor detection efficiencies which, 
in principle, can be arbitrarily reduced. 
A second source of uncertainty is provided by the laws of Quantum Mechanics, and cannot be 
reduced beyond the limits imposed by Heisenberg uncertainty relations
and the Cramer-Rao lower bound \cite{Giovannetti_2006}.
Interferometry with uncorrelated particles allows phase estimations 
with sensitivity bounded by the standard quantum limit (shot-noise) \cite{Pezze_SB} 
\be \label{SN}
\Delta \Theta_{sn} = \frac{1}{N_T^{1/2}},
\ee 
where $N_T$ is the total (or average) number of particles employed in the interferometric process.
Yet, this is not the ultimate limit imposed by Quantum Mechanics.
   
In the last few years, it has become clear that quantum entanglement
has the potential to revolutionize interferometry by allowing phase estimations 
with sensitivities up to the Heisenberg limit \cite{Giovannetti_2006, nota00}
\be \label{HL}
\Delta \Theta_{HL} = \frac{1}{N_T}.
\ee 
Pioneering works along this direction were initiated in the early 80s \cite{Caves_1981}, 
inaugurating a large body of literature proposing new quantum states and strategies for 
sub shot-noise performances \cite{Yurke_1986, Pezze_2006}. 
Quite recently, several efforts have been directed to the experimental realization 
of NOON states \cite{nota_noon, Lamas_2001, Zhao_2004}
(often called Schr\"odinger cats \cite{Leibfried_2003, Leibfried_2004, Leibfried_2005}
in the context of trapped ions):
\be \label{noon}
|\Psi_N \rangle = \frac{1}{\sqrt{2}}\big( |N,0\rangle + e^{i\phi} |0,N\rangle \big).
\ee
The state $|N,0\rangle$ contains $N$
particles in mode $a$ and $0$ particles in mode $b$ (\emph{vice versa} the state $|0,N\rangle$), 
while $\phi$ is an arbitrary phase.
It is widely believed that interferometry with the state Eq.(\ref{noon})  
can estimate unknown phase shifts with sensitivity at
the Heisenberg limit Eq.(\ref{HL})\cite{Lloyd_2004}.
This claim is often accompanied by a simple example.
The phase shift, induced by an external classical perturbing field,
is created by the unitary operator $\hat{U}_{\theta}=e^{-i \hat{J}_z \theta}$, 
where the generator of the unknown phase translation $\theta$
is the two-mode relative number of particles operator, 
$\hat{J}_z = (\hat{N}_a - \hat{N}_b)/2$. 
The projection of the new state 
$|\Psi_N (\theta)\rangle=\hat{U}_\theta |\Psi_N \rangle = (|N,0\rangle + e^{i(\theta N+\phi)} |0,N\rangle)/\sqrt{2}$
over the initial one gives
\be \label{cosN}
|\langle \Psi_{N}|\Psi_{N} (\theta) \rangle|^2 = \cos^2(N\theta/2).
\ee 
Orthogonality, $\langle \Psi_{N}|\Psi_N (\theta) \rangle = 0$, 
is first reached at $\theta =\pm \pi/N$,  which would suggest that the 
smallest measurable phase shift is of the order of $1/N$ as well.
There is a problem, though: in interferometry the incremental phase shift, 
albeit supposedly small, is unknown
and the phase estimation based on the projective measurement Eq.(\ref{cosN})
is ambiguous.
Indeed, $\langle \Psi_{N}|\Psi_N (\theta) \rangle=0$ when  
$\theta = \pm (2 n+1) \pi/N $, with $n =0,1,2,\ldots, N-1$. 
Orthogonality alone is not sufficient to determine $n$, with 
unpleasant consequences when trying to estimate the unknown value of $\theta$ with an arbitrary 
large number of particles and complete prior ignorance.
The $2\pi/N$ oscillation period of Eq.(\ref{cosN}) is typical in  
quantum enhanced technology with state Eq.(\ref{noon}).
For instance, the multipeak structure of Eq.(\ref{cosN}) is present 
(even if not generally discussed) when measuring 
the mean value of the parity operator of one of the output modes obtained after the 
state (\ref{noon}) has been shifted in phase and passed through a 50/50 beam 
splitter \cite{Bollinger_1996, nota01}.
In this Letter we propose i) a measurement strategy for the {\rm unambiguous} estimation of phase shifts
with uncertainty $\sim 1/N_T$ by using the state Eq.(\ref{noon}), within ii) a rigorous Bayesian
analysis of the measurement results which can be implemented experimentally
incorporating decoherence and classical noise and iii) maximum priori
ignorance about the phase shift: $-\pi \leq \theta \leq \pi$
(but we will also consider the case of an arbitrary smaller prior).
The protocol requires $p$ independent interferometric measurements performed with NOON 
states having a different number of particles, $N=1,2,4,...,2^{p-1}$. 
The sensitivity is calculated as a function of the total number of particles used in the
process, $N_T = 2^{p}-1$.

From the experimental point of view, the demonstration of the Heisenberg limit Eq.(\ref{HL})
requires the creation of Schr\"odinger cat states with a minimum of $N=8$
particles, which is within the reach of the present state-of-the-art.
Cat states with up to 6 ions \cite{Leibfried_2005} and 5 photons  \cite{Zhao_2004} 
have been recently reported.
As far as realistic technological applications are concerned, however, Heisenberg limited
interferometry with NOON states  Eq.(\ref{noon}) would likely never overcome the performances of 
classical interferometry Eq.(\ref{SN}), where the
typical number of particles $N_T$ can be several orders of magnitude larger. 
We therefore extend the previous protocol 
to reach {\it sub shot-noise} sensitivity
$\Delta \Theta_{ssn}/\Delta \Theta_{sn} \sim  1/\sqrt{2^{p}-1}$, 
which can be implemented experimentally with an
arbitrarily large number of particles.
We address the possibility to reach over $3$ $db$ sub shot-noise in realistic ion 
and photon experiments within the current technology 
\cite{ Leibfried_2003, Leibfried_2004, Leibfried_2005, Lamas_2001, Zhao_2004},
both in presence of a strong priori constraint and in the more general case of a
complete prior ignorance.
Our results do not only apply to ultra-sensitive interferometry, 
but naturally extend to quantum positioning \cite{Rudolph_2003}, 
clock synchronization \cite{Giovannetti_2001},
frequency standards \cite{Bollinger_1996} and quantum metrology \cite{Giovannetti_2006}.

\begin{figure}[t!]
\begin{center}
\includegraphics[scale=0.55]{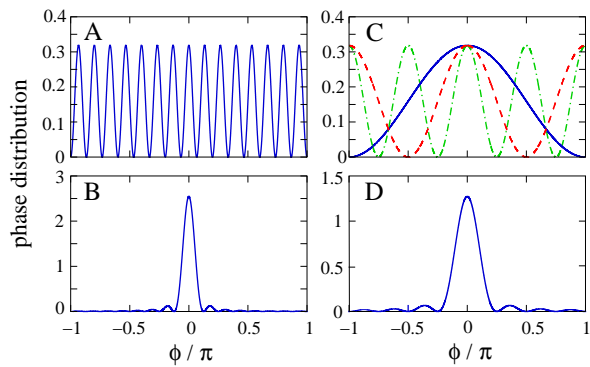}
\end{center}
\caption{ \small{ 
Phase probability distribution Eq.(\ref{prob_distrib}) with $p=1$, $N_T=15$ (A), and
Eq.(\ref{dist2}) with $p=4$, $N_T=15$ (B).
In both cases, the total number of particles is the same, but 
the distribution of B gives a phase sensitivity at the Heisenberg limit.
In C we plot the terms $\cos^2(2^k\phi/2)$ of Eq.(\ref{dist2}). 
The solid blue line is for $k=0$, the dashed red line for $k=1$ and the dot-dashed green line for $k=2$. 
By multiplying these three distributions, as in Eq.(\ref{dist2}), 
all peaks, except the one centered around the 
true value of the phase shift, $\theta=0$, are washed out to give the
phase distribution D ($N_T=7$ and $p=3$). }}\label{proj} 
\end{figure}

\textbf{Bayesian analysis with Schr\"odinger Cats.}
In the following, we discuss the Bayesian phase estimation strategy
considering the Schr\"odinger cat states realized with trapped ions 
by Wineland and collaborators \cite{Leibfried_2003, Leibfried_2004, Leibfried_2005}.
The state $|\Psi_N\rangle=(|N\downarrow\rangle+i^{\xi+N+1}|N\uparrow\rangle)/\sqrt{2}$
was created by applying the ``nonlinear beam splitter'' 
operator $\hat{U}_N=e^{i\frac{\pi \xi}{2}\hat{J}_x}e^{i\frac{\pi}{2}\hat{J}_x^2}$ 
to $N$ spin-down ions 
$|N\downarrow\rangle \equiv |\downarrow\rangle_1...|\downarrow\rangle_N$, 
with $\xi=0$ when $N$ is even, and $\xi=1$ when $N$ is odd \cite{Molmer_1998}.
The $|\Psi_N\rangle$ state is then shifted in phase by an unknown
quantity $\theta$ (which has to be determined)
by applying the operator $e^{-i \theta\hat{J}_z}$. 
The final state, obtained after a further application of $\hat{U}_N$, 
is projected over $|N\uparrow\rangle$.
Quantum Mechanics provides the probability to have $|N\uparrow\rangle$ 
at the output (which will be denoted as a ``yes'' result), given the 
unknown phase shift $\theta$ and the number of particles, $N$, of the cat state: 
$P(\mathrm{yes}|N,\theta) = |\langle \uparrow N|\hat{U}_N e^{-i \theta \hat{J}_z} 
\hat{U}_N|N \downarrow \rangle |^2 = \cos^2 \big(\frac{N\theta}{2} \big)$.
Notice that this probability is identical to Eq.(\ref{cosN}).
The probability to obtain a ``no'' result is, obviously, $P(\mathrm{no}|N, \theta)=1-P(\mathrm{yes}|N,\theta)$.
A single interferometric experiment consists of $p$ independent measurements. 
We collect a number $p_y$ of ``yes'' and $p_n=p-p_y$ of ``no'' results with probability 
$P_p(p_y,p_n|N_T,\theta)$, $N_T$ being the total number of particles used in the $p$ measurements.  
According to the Bayes theorem \cite{Pezze_2006, Helstrom}, we have 
$P_p(\phi|N_T,p_y,p_n)P(N_T,p_y,p_n)=P_p(p_y,p_n|N_T,\phi)P(\phi)$, 
where $P(\phi)$ is fixed by the prior knowledge
and $P(N_T,p_y,p_n)$ provides the normalization of $P_p(\phi|N_T,p_y,p_n)$, which is 
a phase probability distribution. 
We have
\be \label{Pp}
P_p(\phi|N_T,p_y,p_n)\approx \prod_{j=1}^{p_y+p_n} P(\phi|N_j,r_j),
\ee
where $N_T=\sum_{j=1}^{p_y+p_n} N_j$ and $r_j \equiv \mathrm{yes}$ ($\mathrm{no}$)
if, in the $j-{\mathrm{th}}$ measurement, done with $N_j$ particles, we obtain a ``yes'' (``no'') 
result. 
Equation(\ref{Pp}) contains all the available information needed to estimate $\theta$.
We can choose, as estimator $\Theta_{est}$, the maximum of the phase distribution, 
and, as uncertainty $\Delta \Theta$, the $68 \%$-confidence interval \cite{Pezze_2006}, 
namely the phase interval containing $68.27\%$ probability given by
$\int_{\Theta_{est}-\Delta \Theta}^{\Theta_{est}+\Delta \Theta} \ud \phi \, P_p(\phi|N_T,p_y,p_n)=0.6827$.
In the following, for analytical simplicity, we will consider, unless explicitly specified,
the case in which the measurement results are only ``yes'': $p_y = p $, $p_n=0$.
That happens with certainty at $\theta=0$.
The extension to an arbitrary value of $\theta$, where both ``yes'' and ``no''
are accessible, will be discussed in \cite{Pezze}, also 
including decoherence effects.

\textbf{$1/N_T$ periodicity of the Bayesian phase distribution.}
Let us first analyze the interferometric experiment with a state $|\Psi_{N_{T}}\rangle$
of $N_T$ particles and a single measurement: $p=1$.
The phase distribution becomes
\be \label{prob_distrib}
P_1(\phi|N_T,p_y=1, p_n=0) \propto \cos^2( N_T \phi/2).
\ee
This probability contains $N_T$ peaks separated by a distance $2\pi/N_T$, see Fig.(\ref{proj},A). 
Hence, our best guess about the real phase shift $\theta$ is
$\Theta_{est}=\frac{2\pi }{N_T}n \pm \frac{\sqrt{2}}{N_T}$ with $n=0,1, \ldots ,N_T-1$, 
where the error $\sqrt{2}/N_T$ is the mean square fluctuation around a single peak.
In practice, we do not have any alternative but to choose, as phase estimator, 
one of the $N_T$ equivalent peaks of the distribution [cf. discussion after Eq.(\ref{cosN})]. 
In this case, however, the interferometric experiment does not give any substantial 
improvement in phase sensitivity.
Tautologically, the phase uncertainty would scale with the total number of atoms as
$\sim 1/N_T$ only if we have the \emph{a priori} knowledge that the phase lies in an interval of width 
$2 \pi / N_T$ around the real value. 
The $N_T$-peaks structure of Fig.(\ref{proj},A) does not allow the Heisenberg limit, 
even in the case of arbitrary small incremental phase shifts.
How is, therefore, possible to to select the ``right'' peak, so to speak, 
in order to build an unambiguous phase estimator?

\textbf{The $\Delta \theta \sim N^{-3/4}$ limit.}
In this section we consider multiple independent measurements with different 
number of particles.
The first measurement is done with a cat state of a single particle, $N=1$ 
(with a prior knowledge of the phase shift in the interval $[-\pi, \pi]$). 
We then perform a second measurement with $N=2$, and we multiply the resulting distribution 
with the previous one. 
We repeat this procedure $p$ times, increasing, in each shot, the number of particles
in an arithmetic sequence $N=1,2,3,...,p$.
The total number of particles is $N_T=p(p+1)/2$, and the phase distribution is 
\be 
P_p \big( \phi|N_T, p_y=p, p_n=0 \big) \propto \prod_{k=1}^{p}\cos^2\bigg(\frac{k \phi}{2}\bigg).
\ee 
In the limit of large $p$, a Gaussian approximation gives 
$\prod_{k=1}^{p} \cos^2(k\phi) \approx \exp\big[-\frac{\phi^2}{4}\frac{p(p+1)(2p+1)}{6}\big]$, 
where $\sum_{k=1}^{p} k^2= p(p+1)(2p+1)/6$. 
For $p \gg 1$ we have $N_T \sim p^2/2$, and $\Delta \Theta=(9/2)^{1/4}/N^{3/4}_T$, 
which is in good agreement with the numerical calculation:
\be \label{lim34}
\Delta\Theta=\frac{1.44}{N^{3/4}_T}.
\ee 
This argument can be generalized to the case of any prior phase knowledge $[-\pi/L, \pi/L]$, with $L \geq 1$.
The goal is to obtain an unbiased estimate of $\theta$ within a region $2 \pi / L$, the
peaks outside this region being wiped out but the prior knowledge. As before, the protocol requires $p$ measurements:
the first one with $\tilde{N}\approx L$ particles, the  second one  with $2 \tilde{N}$, ...,  
the last one with $p\tilde{N}$. 
The total number of particles is
$\tilde{N} \, p \, (p+1)/2$.
In the limit of large $p$, we obtain an unbiased phase estimate with a sensitivity
$\Delta \Theta= \frac{1.44}{\tilde{N}^{1/4}} \frac{1}{N_T^{3/4}}$.
With an arbitrary value of the true phase shift $\theta \neq 0$, the same protocol provides 
a distribution peaked about $\theta$ with a sensitivity scaling as $\sim 1/N_T^{3/4}$ for $N_T \gg 1$.
This because the $\cos$ and $\sin$ distributions corresponding to ``yes'' and ``no'' results
overlap out of phase and cancel out each other except in a region around $\theta$.
Yet, even if the scaling $\sim 1/N_T^{3/4}$ overcomes the shot-noise Eq.(\ref{SN}),
still this is not the fundamental limit.

\begin{figure}
\begin{center}
\includegraphics[scale=0.33]{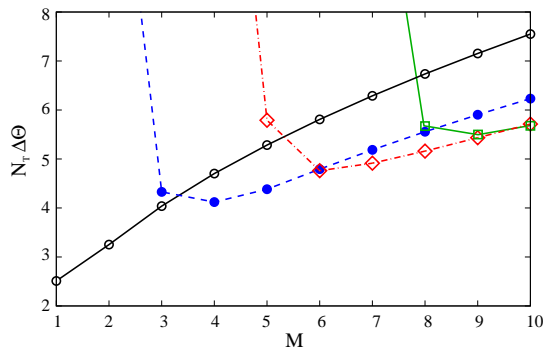}
\end{center}
\caption{\small{Phase sensitivity obtained with $N=1,r,r^2, ..., r^{p-1}$, where 
each measurement (using $N=r^k$ particles) is repeated $M$ times. 
The best strategy is obtained when $r=2$ (black circles) and $M=1$, due to a perfect superposition of maxima and minima 
in the Bayesian distributions, see Fig.(\ref{proj},D).  
Notice that the optimal value of $M$, as well as the phase uncertainty, increases with $r$: here we 
consider $r=3$ (blue points), $r=4$ (red diamonds) and $r=5$ (green squares).
Lines are guide to the eye.}} \label{M} 
\end{figure}

\textbf{The Heisenberg $\Delta \theta \sim N^{-1}$ limit.} 
Let us now consider $p$ independent measurements done with a geometric sequence of 
particles, $N=1,2,4,8,...2^{p-1}$. 
The phase distribution becomes
\be \label{dist2}
P_p(\phi|N_{T}, p_y=p, p_n=0) \propto \prod_{k=0}^{p-1}\cos^2\Big(\frac{2^k \phi}{2}\Big),
\ee
where the total number of particles employed in the complete process is
$N_T=2^{p}-1$.
Fig.(\ref{proj},B) shows the case $p=4$, $N_{T}=15$, with a prior $-\pi \leq \theta \leq \pi$.
Again, the width of the distribution can be simply calculated with a Gaussian
approximation of 
each $\cos^2\big(\frac{2^k \phi}{2}\big)$ term, giving 
$P_p(\phi|N_{T}, p_y=p, p_n=0) \simeq \exp[-\frac{\phi^2}{4} \frac{(2^p+1)(2^p-1)}{3}]$.
In the limit of large $p$ we obtain a phase sensitivity at the Heisenberg limit 
$\Delta \Theta=\sqrt{6}/N_T$.
The numerical calculation gives, 
asymptotically in the number of measurements $p$, 
\be \label{sensHL}
\Delta \Theta=\frac{2.55}{N_T}
\ee 
for a $68\%$ confidence \cite{nota2}.
We therefore conclude that it is possible to obtain an \emph{unbiased} phase 
estimation at the Heisenberg limit,
with repeated measurements and complete prior ignorance.
The trick is to carefully choose the number of particles in each measurement 
and to multiply the corresponding phase probabilities
so to cancel out the extra peaks of the phase distribution, see Figs.(\ref{proj}). 
To clarify this effect, let us consider a phase distribution obtained with $2^k$ 
particles: $\cos^2(2^k\phi/2)$. 
This has maxima in $\phi_{max}^{k}(n)=2\pi n/2^k$, with $n=0, \pm 1, ..., \pm (2^k-1)$.
Conversely, the phase distribution obtained with $2^{k-1}$ particles has 
minima in $\phi_{min}^{k-1}(n)=2 \pi n/2^k$ with $n=\pm 1, \pm 3,  ..., \pm (2^{k-1}-1)$.
By multiplying the two distributions, the maxima $\phi_{max}^{k}(n)$ superimpose with  
the minima $\phi_{min}^{k-1}(n)$, for $n=\pm 1, \pm 3, ... $.
When taking into account also the distributions with $2^{k-2}, 2^{k-3},...$ particles, 
we obtain the cancellation of all peaks $\phi_{max}^{k}(n)$ with $n=\pm1, ...,\pm (2^k-1)$
except the central one, $\phi_{max}^{k}(n=0)$, which is 
enhanced and has a width scaling as $\sim 1/2^{k} \sim 1/N_T$, see Figs.(\ref{proj}, C,D).
This shows that the protocol employing $N=1,2,4,...,2^p$ particles is optimal and 
gives the Heisenberg limit with the best prefactor.   
Indeed, let us consider a general sequence $N=1,r,r^2, ..., r^{p-1}$, with an arbitrary integer $r$.
If $r=1$, we recover the shot-noise limit Eq.(\ref{SN}), 
for $r=2$, we have, as already discussed, a perfect superposition of maxima and minima.
For $r>2$, the phase distribution is characterized by several, strongly weighted, peaks which increase
the phase uncertainty.
To overcome this problem we must repeat $M>1$ times each interferometric measurement with a NOON state
of fixed $N$ before increasing the number of particles (employing a total $N_T=M\sum_{k=0}^{p-1}r^k$).
For a sufficient large $M$, it is possible to recover the Heisenberg limit Eq.(\ref{HL}), but at the price of a 
higher prefactor.
In fact, we have that $\Delta \theta \sim \sqrt{M}/N_T$, as a consequence of the statistics of 
independent measurements.
In Fig.(\ref{M}) we analyzed the sensitivity obtained with different $r$ as a function of $M$: 
for $r=2$ (black circles) the best performance is obtained at $M=1$, for $r=3$ (blue points) at $M=4$,  
for $r=4$ (red diamonds) at $M=6$ and for $r=5$ (green squares) at $M=9$.

As mentioned before, the analysis can be extended to estimate an arbitrary unknown 
phase shift $\theta\neq 0$.
However, in contrast to the $\Delta \theta \sim N_T^{-3/4}$ case, here 
we have to consider \emph{multiple} repeated measurements
in order to concentrate the probability in a single peak, even for the case $r=2$. 
While the scaling $\sim 1/N_T$ is preserved \cite{Pezze}, 
Eq.(\ref{sensHL}) gives a lower bound of phase sensitivity. 

\begin{figure}[t!]
\begin{center}
\includegraphics[scale=0.33]{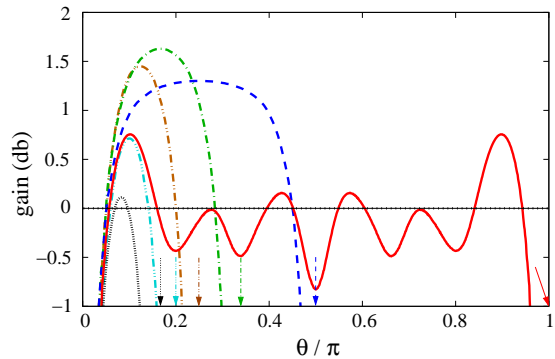}
\end{center}
\caption{\small{
Gain (db) with respect to the shot-noise limit obtained
with the experimental parameters \cite{exp_data}.  
The blue dashed line is the case $N=2$, the green dot-dashed line $N=3$, the brown dot-dot-dashed line $N=4$, 
the sky-blue dot-dot-dot-dashed line $N=5$, and the black dotted line $N=6$. 
Colored arrows indicate the upper bound to the phase prior for each data set.
With maximum priori ignorance, we combine $M$ replica of the Bayesian phase distributions with 
$N=1,2,3,4,5,6$, Eq.(\ref{asymptotic}), (solid red line).
The horizontal dotted line gives the shot-noise. }} \label{gain} 
\end{figure}

\textbf{Sub shot-noise with Beryllium ions.}
The experimental demonstration of the Heisenberg limit would require the
creation of Schr\"odinger cats having $N=8$ particles. 
The biggest Schr\"odinger cat created
experimentally so far is with $N=6$
$ ^{9} \mathrm{Be}^{+}$ ions. 
This is still sufficiently large to reach a {\rm sub shot-noise} phase sensitivity \cite{nota1}. 
In the following, we demonstrate, using the fidelities measured
experimentally in \cite{Leibfried_2003, Leibfried_2004, Leibfried_2005} with Beryllium ions,
the possibility to reach a phase sensitivity gain of 0.8 $db$ with respect to shot-noise for a
priori $-\pi \leq \theta \leq \pi$.
The protocol is quite similar to the one discussed above and requires
Bayesian probabilities calculated with different number of ions
combined with $M$ replica of the measurement process. 
We remark, however, that, in the analysis, we should now replace the ideal 
probability distributions 
with those retrieved experimentally. 
Indeed, we need to include the effects
of noise and decoherence present in the experiments which will
inevitably decrease the sensitivity of the interferometer.
This step can be considered as a ``calibration": once the 
experimental probabilities are retrieved and inverted with Bayes,
the interferometer is ready for its use.
Many sources of experimental imperfections and decoherence 
conspire against phase uncertainty enhancement with cat states. 
Laser intensity fluctuations and magnetic field noise have been discussed in \cite{Leibfried_2005}.
Imperfections in the creation of the state (\ref{noon}) due to non ideal $\hat{U}_N$  
and decoherence have the common effect to decrease the oscillation amplitude of 
$P(\mathrm{yes}|N,\theta)$ \cite{Huelga_1997}. 
\begin{figure}
\begin{center}
\includegraphics[scale=0.66]{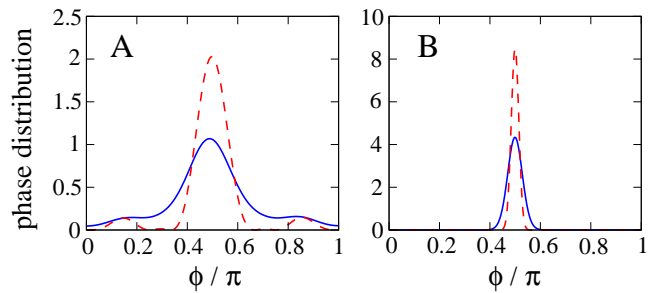}
\end{center}
\caption{\small{Phase probability distributions obtained 
by combining $M$ times (in (A) $M=1$, in (B) $M=10$) the 
Bayesian distributions for $N=1,\, 2,\, 3,\, 4,\,5,\,6$ particles and  $\theta=\pi/2$.
The dashed red line is the ideal case, 
while the solid blue line has been obtained with the experimental fidelities, see \cite{exp_data}.
}} \label{distribution} 
\end{figure}
A fit of the experimental probabilities with
$P_{exp}(\mathrm{yes}|N,\theta) =  A+C_{N \uparrow,N \downarrow}/2 \cos(N\theta)$
and $P_{exp}(\mathrm{no}|N,\theta)  =  1-P_{exp}(\mathrm{yes}|N,\theta)$
has been given in \cite{Leibfried_2003, Leibfried_2004, Leibfried_2005} with
the values of A and $C_{N\uparrow,N\downarrow}$ reported in \cite{exp_data}.
The main effect of the experimental noise is to decrease the fringes contrast
$C_{N \uparrow,N \downarrow}$, which, in the ideal case, is equal to one. 
The experimental Bayesian phase probability distribution associated to a detection of a ``yes''
or ``no" result, $P_{exp}(\phi|N,\{\mathrm{yes}, \mathrm{no}\})$, is obtained inverting $P(\{\mathrm{yes},\mathrm{no}\}|N, \theta)$.
We are now ready to simulate a realistic phase estimation experiment with Beryllium ions: 
i) A ``yes'' or ``no'' result is chosen with probability $P(\{\mathrm{yes},\mathrm{no}\}|N, \theta)$
with an unknown, but fixed, value of $\theta$
and for different number of ions, $N=1,~2,~3,~4,~5,~6$;
ii) We repeat these measurements $M$ times;
iii)  We calculate the Bayesian distribution associated with each ``yes"/``no" result
and values of $N$;
iv) We multiply all Bayesian distributions obtained in iii). This provides
the final phase probability, from which we retrieve
the estimated value of the phase shift and its confidence.
The total number of particles used in this process is $N_T=N_p M$, with
$N_p=\sum_{k=1}^{6}k=21$.
Ideally, with $N=1,2,..., p$, the phase sensitivity would scale as
\begin{eqnarray}
\Delta \Theta= \frac{(9/8)^{1/4}}{N_p^{3/4} M^{1/2}} = \frac{(9/8)^{1/4}}{N_p^{1/4} N_T^{1/2}},
\end{eqnarray}
where $N_p=p(p+1)/2$.
In Fig.(\ref{distribution}) we plot the theoretical and experimental phase probabilities 
for $M=1$ and $M=10$. 
Notice that the experimental distribution for $M=1$ is characterized by a large tail. This 
arises from the reduced fringe visibility and it would 
strongly dilute the confidence of the phase estimation.
On the other hand, by multiplying the $M=10$ distributions, step ii), we strongly 
decrease the weight of the tail with respect to
the central peak, see Fig.(\ref{distribution},B), at the price, of course, to increase
the shot-noise.
It is worth to emphasize, however, that, because of noise and decoherence, 
at the end of the day we could, in principle, get an experimental 
phase sensitivity even worse than the classical shot noise.
Asymptotically in $M$, the phase probability Eq.(\ref{Pp}) can be written as
\begin{eqnarray} \label{asymptotic}
P_{exp}(\phi|N_T,\theta) & = & \prod_{N=1}^6 P_{exp}(\phi|N,\mathrm{yes})^{M  P_{exp}(\mathrm{yes}|N, \theta)} \times \nonumber \\
& & \times P_{exp}(\phi|N,\mathrm{no})^{M  P_{exp}(\mathrm{no}|N, \theta)}.
\end{eqnarray}
In this limit, and with ideal fidelities, we would have a phase independent gain $G_{th}=3.18$ $db$ with
respect to the shot-noise limit $\Delta \Theta_{sn}$.
Such a gain would be comparable with the best performances obtained to date with
photons \cite{McKanzie_2002}.
As a consequence of imperfections and decoherence, 
the experimental gain is lower than the ideal prediction and
depends on the phase shift, see the solid red line in Fig.(\ref{gain}).
The maximum gain is $G_{exp}= 0.75$ $dB$, around the optimal working point
$\theta \sim 0.3$.
In principle, an even higher gain can be obtained
using states with $N=1,2,4,8,...,2^{p-1}$ particles: the 
sensitivity would be bounded by the ideal value 
\be
\Delta \Theta =\frac{2.55}{N_p M^{1/2}}=
\frac{2.55}{N_p^{1/2}N_T^{1/2}}, 
\ee
with $N_p=2^{p}-1$.

So far we have considered the case of complete prior ignorance.
While this can be important for technological and basic science applications like in
gyroscopes or clock synchronizations, it is sometimes possible
to confine the priori within an interval $-\pi/L \leq \theta \leq \pi/L$, with $L > 1$. 
In this case an unbiased phase estimation can be obtained with $M$
replica of the interferometric measurement, each with a
Schr\"odinger cat state of a fixed number of particles
$\tilde{N} \simeq L$.
With ideal distributions, we would obtain a phase-independent sensitivity 
$\Delta \Theta = \frac{1}{\tilde{N}^{1/2}\sqrt{N_T}}$, with $N_T = M~\tilde{N}$. 
The gain with respect to the shot-noise, obtained with the experimental probabilities, 
is shown in Fig.(\ref{gain}) where the arrows indicate the prior knowledge $\pi/\tilde{N}$
for the various cases. 
The gain is maximum at $\theta=\pi/(2 \tilde{N})$ where 
the experimental probabilities to have a ``yes'' or ``no'' result are closer to the ideal ones.
Notice that the sensitivity at first increases and eventually decreases with the number of ions.
This is caused by the competition between the gain obtained increasing $\tilde N$ and the 
loss of visibility due to decoherence, which is higher for larger cats.
With the experimental fidelities measured so far, 
the best scenario is obtained employing cat states of $\tilde{N}=3$ particles, which
provides a gain up to $1.63$ $db$ with respect to the shot-noise.
Yet, it is clear that, with a priori $-\pi/L \leq \theta \leq \pi/L$  
and with cat states having a number of particles $\tilde{N} \simeq 2^k L$, with $k=1,2,3,...$, 
we can, in principle, further decrease the phase uncertainty than using cats with a fixed number of 
particles. 


We remark that the total number of particles used in our phase estimation protocol can be
arbitrarily large (being proportional to the number of replica $M$). 
This is important for realistic technological applications since the number of
particles in Schr\"odinger cat states would probably remain relatively small,
at least in the next future.
We can expect to have, in a few years, the production of robust high-fidelity 
cat states up to $\sim 10$ photons or ions, 
allowing the saturation, at least in principle, of Eq.(\ref{HL}). 
Major obstacles to these efforts are creation imperfections and decoherence \cite{Huelga_1997,zurek}. 
On the other hand, Bose Einstein Condensation might offer the possibility to create 
NOON states with a larger number of particles \cite{Dunningham_2001}. 
A different, promising strategy to experimentally reach sub shot-noise
sensitivities is to use number-squeezed states, which have been 
recently experimentally demonstrated \cite{Orzel_2001,Chuu_2005} with up to 
few hundred neutral atoms. 
In any case, crucial to the success Heisenberg limited interferometry, 
is the realization of highly efficient 
number counting detectors, which can be probably developed in the next generation of
experiments \cite{Chuu_2005, Khoury_2006}.
 
\textbf{Conclusions}. 
Ultra-sensitive interferometry requires unbiased 
phase estimation protocols and carefully engineered maximal quantum 
correlations among input states.
In this paper we have developed a novel Bayesian protocol based on multiple
measurements with Schr\"odinger cat states of variable number of particles. 
This achieves the Heisenberg sensitivity $\Delta \Theta \sim N_T^{-1}$
with an arbitrary prior phase uncertainty.
This protocol overcomes difficulties present in previous approaches
where the estimate was strongly ambiguous, so to require a prior knowledge of the true value of
the phase in the restricted interval $-\pi/N_T \leq \theta \leq \pi/N_T$. 
We have also demonstrated the possibility to obtain sub shot-noise phase sensitivity
with the experimental data published in \cite{Leibfried_2003, Leibfried_2004, Leibfried_2005}
on the creation of Beryllium ions Schr\"odinger cat states.
Our results do not depend on how the state Eq.(\ref{noon}) 
is created, nor on the specific interferometric apparatus as far as 
the phase probability distributions have an oscillating pattern with a period depending on the
number of particles. 

\acknowledgments
We thank John Chiaverini and Jonathan Dowling for useful discussions.


\begin{thebibliography}{0}
\bibitem{nota2} We have $\Delta \Theta^{38\%}=1.25/N_T$ for a $38\%$ confidence
and $\Delta \Theta^{95\%}=14.79/N_T$ for a $95\%$ confidence.
\end{thebibliography}
\end{document}